\documentclass[prd,reprint]{revtex4-2}
\usepackage{float} 
\usepackage{amsmath,amssymb}
\usepackage{graphicx}
\usepackage{hyperref}
\usepackage{color}  
\usepackage[bottom]{footmisc}
\usepackage{times}
\usepackage{newtxmath}
\usepackage{orcidlink}  
\usepackage{natbib}
\hypersetup{   
	colorlinks,
	citecolor=blue,
	linkcolor=blue,
	urlcolor=blue}

\begin{document}
	\title{Effective Potential Approach to the Orbital Motion of Neutral and Charged Particles Around Black Holes}
	\author{Eahsaan Nazir Najar \orcidlink{0009-0006-5959-0887}}
	\email{eahsaan.phy@gmail.com}
    \affiliation{Physics Department, School of Physical and Chemical Sciences, Central University of Kashmir, J\&K-191131, India}
	\date{\today}
	
	\begin{abstract}
    In this paper we study the effective gravitational potential of Kerr, Reissner-Nordström and Kerr-Newman black holes with the relativistic corrections to evaluate the circular orbits and the Innermost Stable Circular Orbits (ISCOs) \textemdash \ a purely relativistic phenomenon \textemdash \ of neutral and charged particles in the vicinity of these spacetimes. We study the circular orbits and ISCOs of black holes mathematically and graphically. Moreover, the astrophysical properties of ISCO are also discussed. We find that the particle entering the ISCO loses a certain amount of energy, in different spacetimes, by gravitational radiative processes, before finally spiraling into ISCO. The electromagnetic effects of different spacetimes show how the charges sharpen the radius of the circular orbits, and the increasing product of particle and black hole charge increases the radius of the ISCO. Most importantly, the effective potential of the most general spacetime, as predicted by the no-hair conjecture, has been derived.

\end{abstract}

	\maketitle
	
	\section{Introduction}
Among all the physically plausible causal structures, black holes are the most fascinating and relativistically interesting ones. We study black holes through their interactions with surrounding matter or via their spacetime deformations outside the event horizons: the causal boundary between events that can communicate with distant observers and events that cannot \cite{MTW1973,Schutz2009, Wald1984}. Black holes are interesting for being one of the possible end-points of stellar evolution, they are the power feeders of quasars and active galactic nuclei and are, nevertheless, the most mysterious macroscopic objects in the universe. It is possible to study them in a way, with mathematical rigor, that is almost impossible for any other stellar system; our interest, however, is in relativistic kinematic effects. \par
In this article we explore the circular orbits and ISCOs of neutral and charged particles in the gravitational field of Kerr, Reissner-Nordström, and Kerr-Newman black holes. ISCOs represent the region between test particles orbiting the black hole and particles falling into the black  hole; therefore, they are an important tool for studying the physics of accretion disk \cite{Shakura1973,Jaia,Sakia2021}. The geometry of the Schwarzschild spacetime has been studied by \cite{Romero2014} and the circular orbits and ISCOs of neutral particles in the Schwarzschild black hole, where the black hole has been assumed to be charge neutral, by \cite{AlZehrani2021}. Black holes are widely considered charge neutral as shortly after the accretion of ambient matter neutralizes all excess charges. But we have strong reasons to believe that weakly charged black holes exist \cite{Carter1973,Zajacek2019a,Zajacek2019b,Zajacek2018} and the electromagnetic coupling of the black hole charge and test particle charge takes place. \par
The Schwarzschild black holes are non rotating, so, following the same procedure  we study the Kerr black holes; section~\ref{sec:kerr}. The circular orbits and ISCOs of neutral particles in the vicinity of Kerr spacetime are evaluated in section~\ref{cogkerr}. The problem of interaction of black hole charge and test particle charge in the Kerr spacetime is the subject matter of section~\ref{iscokerr}; here we have studied the ISCOs of charged particles immersed in an axisymmetric magnetic field \cite{Wald1974,AlZehrani2021}.\par
Next we study Reissner-Nordström,  another important class of back holes, section~\ref{sec:rnbh}; followed by the study of circular orbits and ISCOs of neutral particles in the vicinity of Reissner-Nordström spacetime, section~\ref{cogrnbh} and ISCOs of charged particles in Reissner-Nordström black hole spacetime~\ref{iscornbh}. We calculated the effective potential of the Reissner-Nordström spacetime.\par
Finally, we investigated the problem of Kerr-Newman black hole \cite{Sakia2021,Schroven2017,Hackmann2013}: geometry, circular orbits, and ISCOs of neutral particles, respectively, in sections~\ref{knbh} and~\ref{cogknbh}. We have derived the most effective gravitational potential of a black hole as stated by the no-hair conjecture \textemdash \ a stationary black hole parametrization is done solely by three parameters: mass, charge, and angular Momentum \textemdash \ because in Kerr-Newmann spacetimes, contrary to Schwarzschild, Kerr, and Reissner-Nordstrom spacetimes, all three parameters, stated above, are nonzero. Thus, the Kerr-Newmann black hole according to the no-hair theorem is the most general stationary solution to Einstein's field equations. \par
We use the sign conventions adopted by \cite{MTW1973} and the speed of light($c$), the universal gravitational constant($G$) and the Coulomb's constant($k$) are unity.
    \section{Kerr Black Hole}\label{sec:kerr}
The Schwarzschild black hole is non-rotating, a heuristic solution of Einstein's field equation for rotating body of mass $M$ and specific angular momentum $a$  was derived by R. Kerr \cite{Kerr_1963}. The Kerr metric is the most unique stationary axisymmetric vacuum solution (Carter-Robinson theorem), just as the Schwarzschild metric is the most static vacuum solution (Israel’s theorem). \par
In Boyer Lindquist coordinates
\begin{equation}
\Delta=r^2-2Mr+a^2 \ \ \text{and} \ \  \rho^2= r^2+a^2\cos^2\theta,
\end{equation}
since ${g}_{\mu \nu }(r,\theta)$ is symmetric, so ${g}_{t\phi} = {g}_{\phi t}$. The squared line element of the Kerr metric is:
\begin{eqnarray}
    ds^2&=&-\bigg(\frac{\Delta - a^2{\sin}^2\theta}{\rho^2}\bigg)dt^2 -\frac{4aMr\sin^2\theta}{\rho^2}dtd\phi \nonumber \\ &+& \frac{\Gamma\sin^2\theta}{\rho^2}d\phi^2 +\frac{\rho^2}{\Delta}dr^2 + \rho^2d\theta^2
\end{eqnarray} 
We will study the circular orbits of neutral particles and the innermost stable circular orbits of neutral and charged particles in the Kerr field.
\subsection{Circular orbits of gravitating neutral particles in Kerr field}\label{cogkerr}
In this section, we discuss how particles orbit around the Kerr black hole in the absence of electric and magnetic fields. An interesting sub-manifold to evaluate the quadrature is the equatorial sub-manifold. For an equatorial sub-manifold:
\begin{center} 
$\theta=\frac{\pi}{2}, \ \ \dot{\theta}=0, \ \ \rho^2=r^2 \ \ \text{and} \ \ \Gamma=(r^2+a^2)^2-a^2\Delta$.
\end{center} 
Hereinafter $\Gamma$ is the Boyer Lindquist coordinate for the equatorial manifold. The squared line element of the Kerr spacetime in the equatorial sub-manifold is
\begin{equation}
    ds^2=-\bigg(\frac{\Delta - a^2}{r^2}\bigg)dt^2 -\frac{4aM}{r^2}dtd\phi + \frac{\Gamma}{r^2}d\phi^2 +\frac{r^2}{\Delta}dr^2 + r^2d\theta^2.
\end{equation}
The Kerr metric, just like the Schwarzschild metric, due to its temporal and azimuthal symmetry, allows two killing vectors
\begin{eqnarray}\label{sidd}
\psi^{\mu} = \bigg(\frac{\partial}{\partial t}\bigg)^{\mu} \quad \text{and} \quad  \chi^{\mu} = \bigg(\frac{\partial}{\partial \phi}\bigg)^{\mu}.
\end{eqnarray}
The corresponding conserved quantities are the energy and the angular momentum, respectively. If the four-velocity of a test particle in the Kerr background is $u^{\mu} $, then, the contraction of four-velocity is $g_{ab}u^au^b= -k$  with $k=1 $ for timelike events and $0 $ for lightlike events.
\par
The specific energy and specific angular momentum take the form:
\begin{equation}
   \mathcal{E}=-u^{\mu}\psi_{\mu} =-g_{\mu \nu}u^{\mu}u^{\nu}  \quad \text{and} \quad \mathcal{L}=u^{\mu}\chi_{\mu}.
\end{equation}
The component-wise and explicit forms of the specific energy and specific angular momentum are
\begin{equation}\label{39}   
\mathcal{E}=\bigg(\frac{\Delta-a^2}{r^2}\bigg)\dot{t} + \frac{2Ma}{r}\dot{\phi} \quad \text{and} \quad \mathcal{L}=\frac{-2Ma}{r}\dot{t}+ \frac{\Gamma}{r^2}\dot{\phi}.
\end{equation}
The expanded form of $g_{ab}u^au^b= -k$ is 
\begin{equation}\label{40}
    \bigg(\frac{\Delta-a^2}{2}\bigg)\dot{t}^2 +\frac{4aM}{r}\dot{t}\dot{\phi} -\frac{\Gamma}{r^2}\dot{\phi}^2-\frac{r^2}{\Delta}\dot{r}^2 =k.
\end{equation}
The radial equation of the equatorial sub-manifold can be written as the function of $\mathcal{E}$ and $\mathcal{L}$ instead of $\dot{t}$ and $\dot{\phi}$, using equations \eqref{39} to write $\dot{t}$ and $\dot{\phi}$ as functions of $\mathcal{E}$ and $\mathcal{L}$, we get
\begin{equation}\label{41}
\dot{t}={r^2 \over \Delta -a^2} \bigg[\mathcal{E}-{2Mar\mathcal{L}(\Delta -a^2)+4M^2a^2r^2\mathcal{E} \over \Gamma(\Delta -a^2)+M^2a^2r^2}\bigg]
\end{equation}
and 
\begin{equation}\label{42}
\dot{\phi}= \bigg[\frac{\mathcal{L}(\Delta-a^2)r^2+2Mar^3\mathcal{E}}{\Gamma(\Delta-a^2)+4a^2M^2r^2}\bigg].
\end{equation}\\
Now, substituting \eqref{41} and \eqref{42} in \eqref{40}, we get
\begin{widetext}
\begin{eqnarray}\nonumber\label{43}
\dot{r}^2 &=& \Bigg\{-k+ \bigg[ \frac{r^2}{\Delta-a^2}\bigg] \bigg[\mathcal{E}-\frac{\mathcal{L}(\Delta-a^2)2Mar+4M^2r^2a^2\mathcal{E}}{\Gamma(\Delta-a^2)+4M^2a^2r^2}\bigg]^2 -\frac{\Gamma}{r^2}\bigg[\frac{\mathcal{L}(\Delta-a^2)r^2+2Mar^3\mathcal{E}}{\Gamma(\Delta-a^2)+4a^2M^2r^2}\bigg]^2\\\nonumber\\&+&\frac{4aMr}{(\Delta-a^2)}\bigg[\mathcal{E}-\frac{\mathcal{L}(\Delta-a^2)2Mar+4M^2r^2a^2\mathcal{E}}{\Gamma(\Delta -a^2)+4M^2a^2r^2}\bigg]\bigg[\frac{\mathcal{L}(\Delta-a^2)r^2+2Mar^3\mathcal{E}}{\Gamma(\Delta-a^2)+4M62a^2r^2}\bigg]\Bigg\}\Bigg/\frac{r^2}{\Delta}.
\end{eqnarray}
\end{widetext}
An immediate verification of this radial equation can be performed using the fact that when the Kerr parameter $a$ is set to $0$, the Kerr metric reduces to the Schwarzschild metric. Therefore, $\Delta=r^2-2Mr$ and  $\Gamma=r^4 $ and this reparametrization for timelike geodesics $(k=1)$ reduce the equation \eqref{43} into
 \begin{equation}
    \dot{r}^2 + \bigg(1-\frac{2M}{r}\bigg) \bigg(1+\frac{\mathcal{L}^2}{r^2}\bigg)=\mathcal{E}^2,
\end{equation}
which is same as for the Schwarzschild black hole, see equation 6.3.14 of \cite{Wald1984}. The radial dynamical equation \eqref{43}, therefore, is more general and applicable to all the test particles outside the ergo-sphere of the Kerr black hole. This equation also assumes two non extraneous solutions, i.e., positive and negative solutions. The analytic solution of this differential equation is very difficult. After a lot of algebraic work, we found the solution of \eqref{43} for a timelike geodesic as 
\begin{widetext}
\begin{equation}\label{45}
\dot{r}_{\pm}=\pm\sqrt{\frac{\mathcal{E}^2 \left(a^2 (2 M+r)+r^3\right)-a^2 r-4 a \mathcal{E}  \mathcal{L} M+\left(\mathcal{L}^2+r^2\right) (2 M-r)}{r^3}}.
\end{equation}
\end{widetext}
We can validate equation \eqref{43} simply by squaring equation \eqref{45} and setting $a=0$, which reduces equation \eqref{45} to equation 7.163 of \cite{TPad}, as expected.\par
\begin{figure}[h]
	\centering
\includegraphics[scale=0.9]{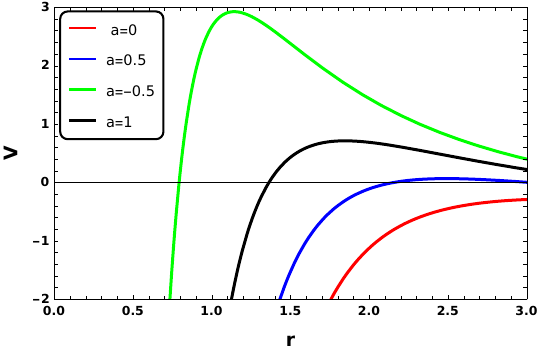} \ \ \includegraphics[scale=0.9]{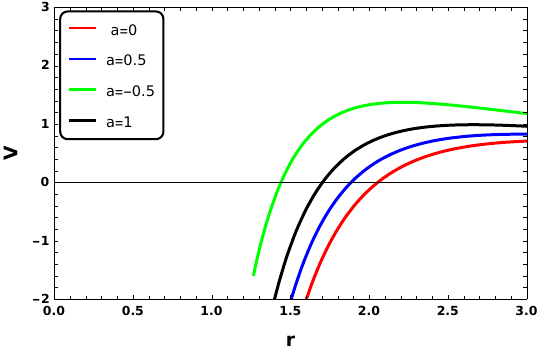}
\caption{\small Variation of effective potential of Kerr black hole with $r$ for different values of $a$ with $\mathcal{E}=0.5$ and (Top) and $\mathcal{E}=1$ (Bottom).}
\label{v}
\end{figure}

For the null geodesics (or lightlike events for which $k=0$), the solution takes the form of
\begin{equation}\label{46}
\dot{r}_{\pm}=\pm\sqrt{\frac{\mathcal{E}^2 \left(a^2 (2 M+r)+r^3\right)-4 a \mathcal{E} \mathcal{L} M+\mathcal{L}^2 (2 M-r)}{r^3}}.
\end{equation}
By squaring equations \eqref{46} and setting $a=0$, the result turns out to be Equation 6.3.29 of Wald \cite{Wald1984}, which reads the effective potential as
\begin{equation}
V_{\it{eff}}={\mathcal{L}^2 \over 2r^3}(r-2M).
\end{equation}
Just like the Schwarzschild metric, Kerr black holes also show the phenomenon of innermost stable circular orbits. We can convert equation \eqref{45} into classical quadrature by squaring
\begin{eqnarray}\nonumber
\dot{r}^2r^3=\mathcal{E}^2 \left(a^2 (2 M+r)+r^3\right)-a^2 r- \qquad \qquad\\ 4 a \mathcal{E}  \mathcal{L} M+  \left(\mathcal{L}^2+r^2\right) (2 M-r),
\end{eqnarray}
and upon simplification, the quadrature of a Kerr metric in an equatorial sub-manifold takes the form as
\begin{equation}
{1 \over 2}\dot{r}^2-{M \over r}+{\mathcal{L}^2 \over 2r^2}+{1 \over 2}(1-\mathcal{E}^2)\bigg(1+{a^2 \over r^2}\bigg)-{M \over r^3}(\mathcal{L}-a\mathcal{E})^2=0.
\end{equation}
The relativistically effective potential, thus, is given as
\begin{equation}\label{tata}
V_{\it{eff}}(r,\mathcal{E}, \mathcal{L})=-{M \over r}+{\mathcal{L}^2 \over 2r^2}+{1 \over 2}(1-\mathcal{E}^2)\bigg(1+{a^2 \over r^2}\bigg)-{M \over r^3}(\mathcal{L}-a\mathcal{E})^2.
\end{equation}
The variation of the effective potential of the Kerr black hole with respect to $r$ for different values of charge $a$ is shown in figure \ref{v}.
\par
The condition for any test particles to execute circular motion in the Kerr field is the simultaneous solution of $V_{\it{eff}} = 0$ and $V'_{\it{eff}} = 0$. Now, as
\begin{equation}\label{BYBY}
V'_{\it{eff}} = -\frac{a^2 \left(1-\mathcal{E} ^2\right)}{r^3}+\frac{3 M (\mathcal{L}-a \mathcal{E} )^2}{r^4}-\frac{\mathcal{L}^2}{r^3}+\frac{M}{r^2}=0,
\end{equation}
and the roots of this equation are
\begin{eqnarray}\label{50}
r_{\pm}&=&{\mathcal{L}^2+a^2-a^2\mathcal{E}^2 \over 2M} \nonumber \\ &\pm& {\sqrt{(a^2\mathcal{E}^2-a^2-\mathcal{L}^2)^2-12M^2(\mathcal{L}-a\mathcal{E})^2} \over 2M}.
\end{eqnarray}
Now, if $(a^2\mathcal{E}^2-a^2-\mathcal{L}^2)^2 < 12M^2(\mathcal{L}-a\mathcal{E})^2$  then there are no maxima or minima of $V(r,\mathcal{E}, \mathcal{L})$ which means the particle is trapped and the information is lost. However, the physically most important case is when $(a^2\mathcal{E}^2-a^2-\mathcal{L}^2)^2 > 12M^2(\mathcal{L}-a\mathcal{E})^2$. It has been verified that $r_{+}$ is a minima of $V(r,\mathcal{E}, \mathcal{L})$ and $r_{-}$ is a maxima. The use of equation \eqref{50} in equations \eqref{tata} and \eqref{BYBY} produces the expressions for specific energy and specific angular momentum, respectively, as 
\begin{equation}\label{Ein}
\mathcal{E}_{0}={(r_{0}-M)r_{0}+(r_{0}^2+a^2-2Mr_{0}) \over 2r_{0}\sqrt{r_{0}^2+a^2-2Mr_{0}}}
\end{equation}
and 
\begin{equation}\label{Tein}
\mathcal{L}_{0}={r_{0}^2+a^2 \over a}\mathcal{E}_{0}-{a\sqrt{r_{0}^2+a^2-2Mr_{0}} \over r_{0}}.
\end{equation}
Now the condition for Innermost stable circular orbits to exist necessitates the second derivative of effective potential \eqref{tata} to vanish, that is,
\begin{eqnarray}\label{CON}
a^2 \left(\mathcal{E}_{0}^2 (3 M+r_{0})-r_{0}\right)-6 a \mathcal{L}_{0} M \mathcal{E}_{0} \qquad \nonumber \\ +\mathcal{L}_{0}^2 (3 M-r_{0})+M r_{0}^2=0.
\end{eqnarray}
The sufficient and necessary condition for the radius of orbit to be equal to $r_{\mathcal{ISCO}}$ is algebraically obtained by putting equations \eqref{Ein} and \eqref{Tein} in equation \eqref{CON}, which gives
\begin{equation}
r_{\mathcal{ISCO}}(r_{\mathcal{ISCO}}-6M)+8a\sqrt{Mr_{\mathcal{ISCO}}}-3a^2=0.
\end{equation}
The energy and angular momentum of a test particle in an ISCO of Kerr black hole are respectively
\begin{equation}\nonumber
\mathcal{E}_{\mathcal{ISCO}}=\sqrt{1-{2M \over 3r_{\mathcal{ISCO}}}}
\end{equation}
and
\begin{equation}\label{Tho}
\mathcal{L}_{\mathcal{ISCO}}=\sqrt{{2M \over 3r_{\mathcal{ISCO}}}(3r_{\mathcal{ISCO}}^2-a^2)}.
\end{equation}
Here, it should be noted that $r_{\mathcal{ISCO}}$ lies in the interval $[M,9M]$. The value of $\mathcal{E}_{\mathcal{ISCO}}$ when $r_{\mathcal{ISCO}}=2M$ is $\sqrt{2}/\sqrt{3}$. A particle entering the ISCO will release an energy of up to $1-\sqrt{2}/\sqrt{3}\approx18.3\%$ of its rest energy. 
\subsection{ISCO of charged particles in Kerr black hole}\label{iscokerr}
After discussing the relativistic properties of neutral particles in the Kerr field, we are in a position to discuss the electrodynamic properties of charged particles in the Kerr field. As already stated in equation \eqref{sidd}, the Kerr metric allows for two symmetries \textemdash temporal and azimuthal. We can use these two symmetries to construct the relativistic electromagnetic four potential:
\begin{equation}
A^{\mu}=\Big(aB-{Q \over 2M}\Big)\psi_{(t)}^{\ \mu}+{B \over 2}\psi_{(\phi)}^{\ \mu},
\end{equation}
around a charged Kerr black hole immersed in an axisymmetric magnetic field, see \cite{Wald1974}. Here, $Q$ is the black hole charge and $B$ is the axisymmetric magnetic field strength. Now, we find the specific energy and specific angular momentum of a charged particle in the Kerr field, respectively, as
\begin{eqnarray}
\mathcal{E}&=&-{p_{\mu}\psi^{\mu}_{(t)} \over m} \nonumber \\ \nonumber \\ &=& {q-4Mab \over r}+\Big(1-{2M \over r}\Big)\dot{t}+{2Ma \over r}(b + \dot{\phi})
\end{eqnarray}
and
\begin{eqnarray}
\mathcal{L}&=&{p_{\mu}\psi^{\mu}_{(\phi)} \over m}=\Bigg[{a(q-4Mab) \over r} \nonumber \\ &-&{2Ma \over r}\dot{t}+{r^3+a^2r-2Ma^2 \over r}(b+\dot{\phi})\Bigg].
\end{eqnarray}
Here, we have parameterized $q=eQ/m$ and $b=eB/2m$. We can construct the radial quadrature of the Kerr field as
\begin{widetext}
\begin{eqnarray}
r^3\dot{r}^2=\Bigg({q-2abM \over r} &-&\mathcal{E}\Bigg)^2\Big\{r^3 +ra^2-2Ma^2\Big\}+4Ma\Bigg({q-2abM \over r} -\mathcal{E} \Bigg) \times \Bigg[{(q-4abM)a \over r} - b\Bigg\{{r^3 +ra^2-2Ma^2 \over r}\Bigg\}+\mathcal{L}\Bigg] \nonumber \\ &-& r\Bigg[\Bigg(1-{2M \over r} \Bigg)\label{Garg}  \times \Bigg\{r^2+\Bigg({(q-4abM)a \over r} - b{r^3 +ra^2-2Ma^2) \over r}+\mathcal{L}\Bigg)^2\Bigg\}+a^2\Bigg].
\end{eqnarray}
\end{widetext}
The weak field approximation for Kerr black holes also breaks down when the electric and magnetic fields near black holes create curvature comparable to the black holes mass near the event horizon, see \cite{AlZehrani2021}. Due to the complexity of equation \eqref{Garg}, we cannot write the explicit expressions for the first and second derivatives of equation \eqref{Garg}. We will study the behavior of charges around weakly charged and weakly magnetized black holes graphically, rather than algebraically.
\begin{figure}[h]
\includegraphics[scale=0.832]{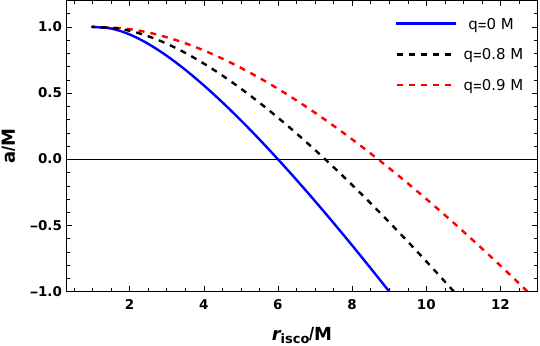} \ \ \includegraphics[scale=0.832]{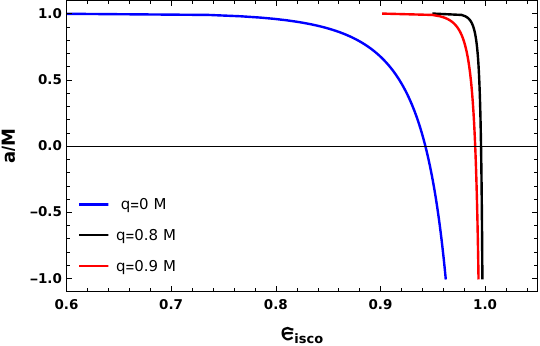} \ \ \includegraphics[scale=0.832]{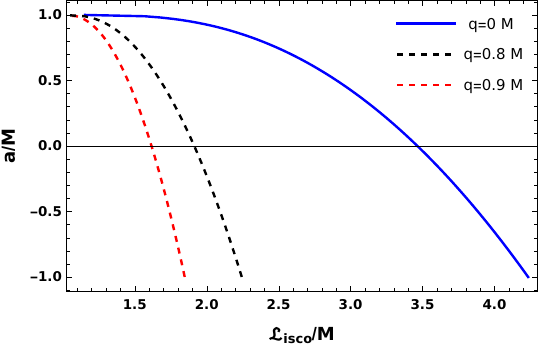}
\caption{\small Variation of radius $r_{\mathcal{ISCO}}/M$ (Top), energy $\mathcal{E}_{\mathcal{ISCO}}$ (Center) and angular momentum $\mathcal{L}_{\mathcal{ISCO}}/M$ (Bottom) with  $a/M$ for  positive values of charge parameter $q$.}
\label{va}
\end{figure}

We will first explain the electric and magnetic effects separately. If we switch off the magnetic field, i.e., $b=0$, the problem is still extremely complex to solve by any analytical method. We have used numerical simulations to study the phenomenon graphically.
\begin{figure}[h]
	\centering
\includegraphics[scale=0.832]{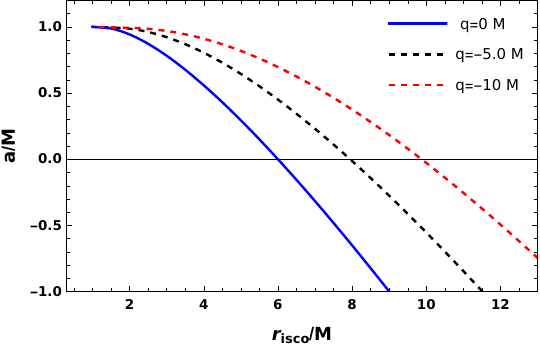} \ \ \includegraphics[scale=0.832]{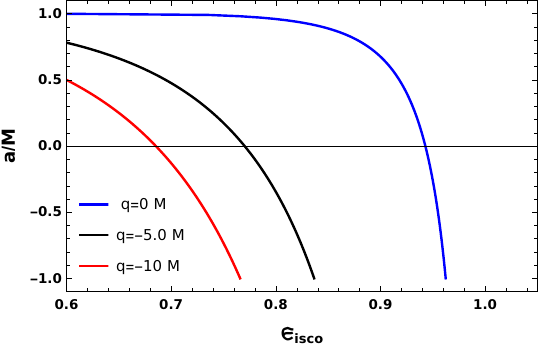} \ \ \includegraphics[scale=0.832]{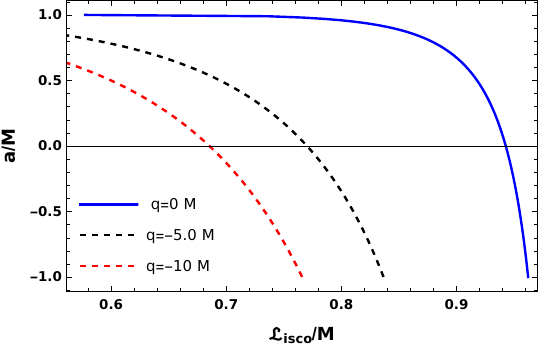}
\caption{\small Variation of radius $r_{\mathcal{ISCO}}/M$ (Top), energy $\mathcal{E}_{\mathcal{ISCO}}$ (Center) and angular momentum $\mathcal{L}_{\mathcal{ISCO}}/M$ (Bottom) with  $a/M$ for negative values of charge parameter $q$.}
\label{var}
\end{figure}

\begin{figure}[h]
	\centering
\includegraphics[scale=0.832]{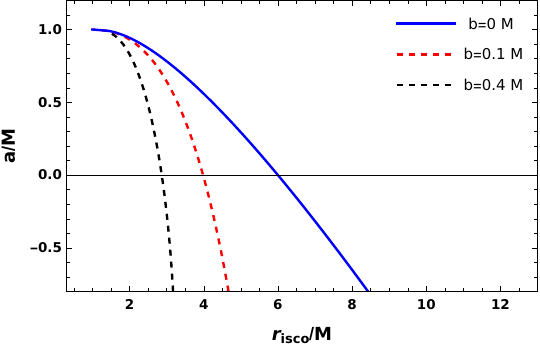} \ \ \includegraphics[scale=0.832]{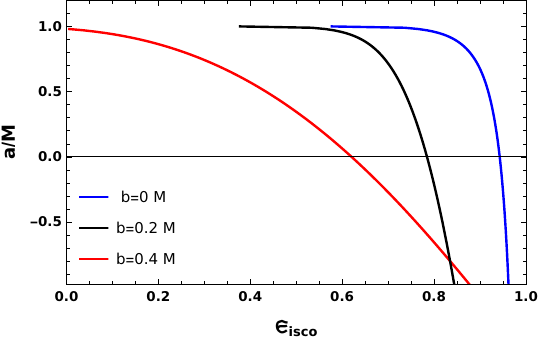} \ \ \includegraphics[scale=0.832]{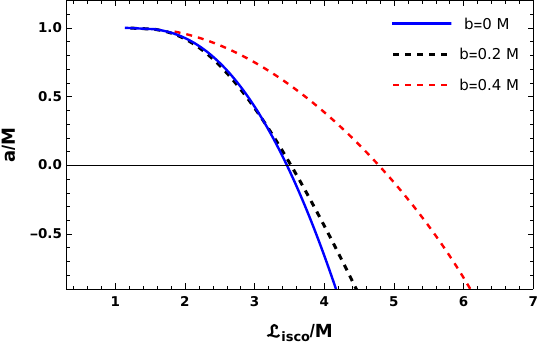}
\caption{\small Variation of  $r_{\mathcal{ISCO}}/M$ (Top),$\mathcal{E}_{\mathcal{ISCO}}$ (Center) and  $\mathcal{L}_{\mathcal{ISCO}}/M$ (Bottom) with  Kerr parameter $a/M$ for positive values of magnetic parameter $b$.}
\label{vari}
\end{figure}
\begin{figure}[h]
	\centering
\includegraphics[scale=0.832]{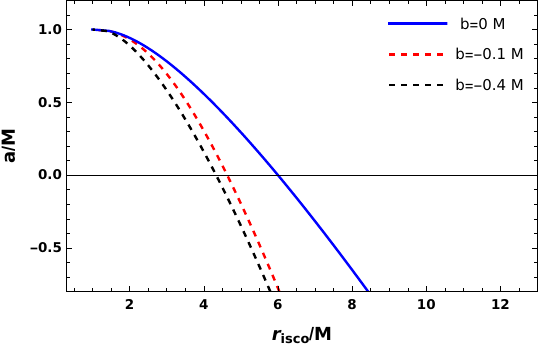} \ \ \includegraphics[scale=0.832]{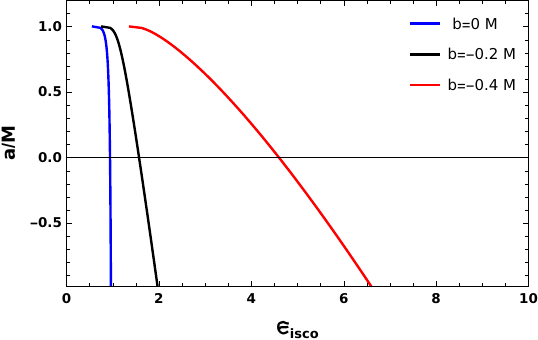} \ \ \includegraphics[scale=0.832]{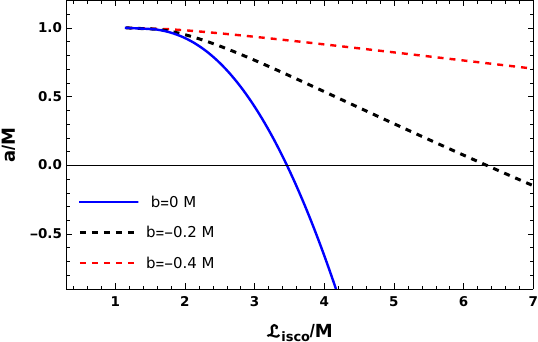}
\caption{\small Variation of  $r_{\mathcal{ISCO}}/M$ (Top),$\mathcal{E}_{\mathcal{ISCO}}$ (Center) and $\mathcal{L}_{\mathcal{ISCO}}/M$ (Bottom) with  Kerr parameter $a/M$ for negative values of magnetic parameter $b$.}
\label{varia}
\end{figure}
The graphs shown above (Figure \ref{va}) were plotted for the positive values of the charge parameter, which means when the test particle and the black hole have the same charge. Again we will plot the graphs (Figure \ref{var}) for the negative values of the charge parameter i.e., when the test particle and the black hole are of opposite charge.

\begin{figure}[h]
\includegraphics[scale=0.832]{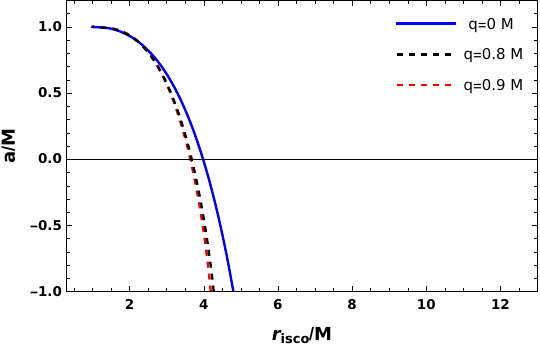} \ \ \includegraphics[scale=0.832]{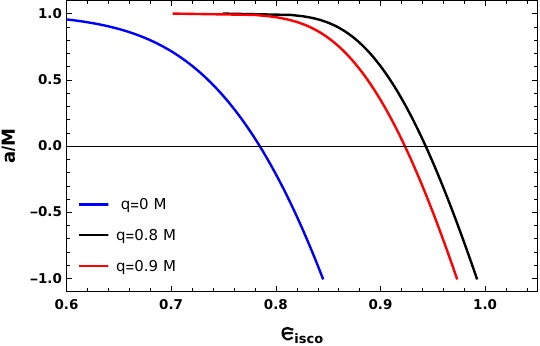} \ \ \includegraphics[scale=0.832]{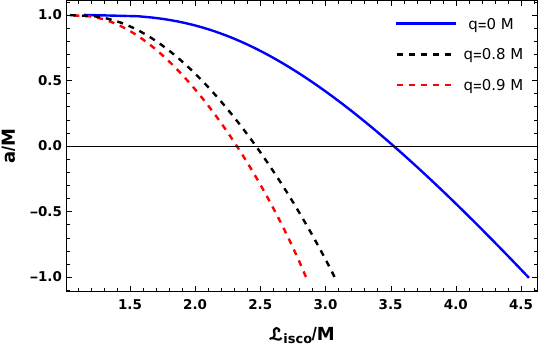}
\caption{Variation of  $r_{\mathcal{ISCO}}/M$ (top),$\mathcal{E}_{\mathcal{ISCO}}$ (Center) and  $\mathcal{L}_{\mathcal{ISCO}}/M$ (bottom) with Kerr parameter $a/M$ for positive values of charge parameter $q$ with $b=0.1/M$.}
\label{variat}
\end{figure}
\begin{figure}[h]
\includegraphics[scale=0.832]{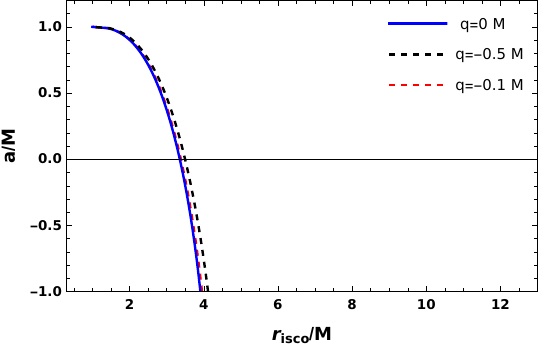} \ \ \includegraphics[scale=0.832]{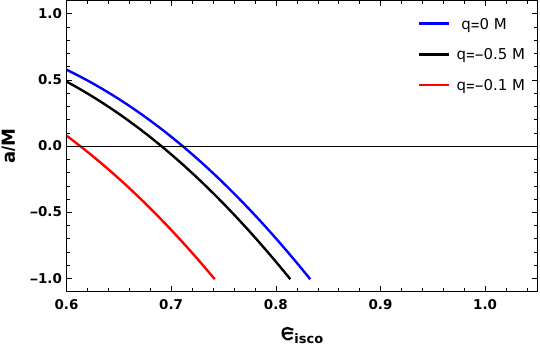} \ \ \includegraphics[scale=0.832]{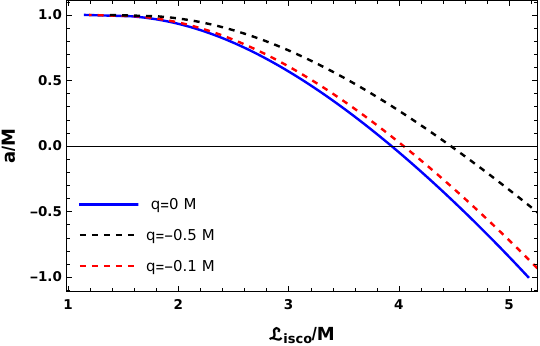}
\caption{\small Variation of  $r_{\mathcal{ISCO}}/M$ (top),$\mathcal{E}_{\mathcal{ISCO}}$ (Center) and $\mathcal{L}_{\mathcal{ISCO}}/M$ (bottom) with Kerr parameter $a/M$ for  negative values of charge parameter $q$ with $b=0.2/M$.}
\label{variati}
\end{figure}
\par
From both figures it can be readily conceived that with an increase in the magnitude of the charge parameter $q$ the radius of ISCO increases. An enormously important point that we obtained is that when $q=M$ then the ISCO falls in the interval $[M,\infty]$ with $\mathcal{E}_{\mathcal{ISCO}}=1$ and $\mathcal{L}_{\mathcal{ISCO}}=M$ for all values of $r$. This perplexing conclusion tells us that a charged particle at infinity can be entangled to a Kerr black hole for which $q=M$. Next, we study the variation of radius $r_{\mathcal{ISCO}}/M$, energy $\mathcal{E}_{\mathcal{ISCO}}$ and angular momentum $\mathcal{L}_{\mathcal{ISCO}}/M$ with the Kerr parameter $a/M$ for different values of the magnetic parameter $b$ by switching off the charge parameter $q$, i.e., $q=0$.

The graphs shown above (Figure \ref{vari}) were plotted for the positive values of the magnetic parameter, which means when the charge of a test particle and the magnetic field of the black hole have the same sign. Now, we plot the graphs (Figure \ref{varia}) for the negative values of the magnetic parameter i.e., when the charge of a test particle and the magnetic field of the black hole are of opposite sign.
\par
Lastly, we can check the variation of radius $r_{\mathcal{ISCO}}/M$, energy $\mathcal{E}_{\mathcal{ISCO}}$ and angular momentum $\mathcal{L}_{\mathcal{ISCO}}/M$ with the Kerr parameter $a/M$ for different values of the magnetic parameter by switching on the magnetic parameter $b$, i.e., $b\neq0$.
\par
The graphs shown above (Figure \ref{variat}) were plotted for the positive values of the charge parameter in the presence of a magnetic field, which means when the test particle and the black hole have the same charge. Again, we plot the graphs (Figure \ref{variati}) for the negative values of the charge parameter in presence of magnetic field, i.e., when the test particle and the black hole are of opposite charge.

    \section{Reissner-Nordström black hole}\label{sec:rnbh}
The next important class of black holes is the famous Reissner-Nordström. These are the spherically symmetric non-rotating charged black holes. The spin parameter for these black holes vanishes, i.e., $a=0$, but they have the charge, say $e$. The geometry of Reissner-Nordström suggests that the true physical singularity no longer has a ring structure. The metric of the Reissner-Nordström black hole in the spherical polar coordinates has the form \cite{Reissner1916,Nordstrom1918}:
\begin{equation}
	ds^2=-{\Delta \over r^2}dt^2 +{r^2 \over \Delta}dr^2+r^2d\theta^2+r^2\sin^2\theta d\phi^2.
\end{equation}
We can make a Boyer-Lindquist-like substitution to enhance the reduction:$ \  \Delta = r^2 +e^2 -2Mr$. As long as the charge distribution of Reissner-Nordström black hole is spherically symmetric, Birkhoff's theorem is applicable. The physical boundaries of the metric are determined by setting $\Delta=0$, which implies
\begin{equation}\nonumber
r_{\pm}=M \pm \sqrt{M^2-e^2}.
\end{equation}
The event horizons for $r_{\pm}$ exist only when $M^2\geqslant e^2$. Throughout the discussion of Reissner-Nordström black holes, we will assume this condition to hold.
\subsection{Circular orbits of gravitating neutral particles in Reissner-Nordström black hole}\label{cogrnbh}
In this section we discuss the circular orbits of neutral particles in the equatorial sub-manifold of black hole. We know, in an equatorial region: $\theta={\pi \over 2}$ and $\dot{\theta}=0$. The Reissner-Nordström metric, ditto as Kerr metric, due to its temporal and azimuthal symmetry, allows two killing vectors
\begin{equation}\nonumber
\psi^{\mu} = \bigg(\frac{\partial}{\partial t}\bigg)^{\mu} \ \quad \ \text{and} \ \quad \  \chi^{\mu} = \bigg(\frac{\partial}{\partial \phi}\bigg)^{\mu}.
\end{equation}
The corresponding conserved quantities are the energy and the angular momentum, respectively. If the four-velocity of a test particle in the gravitational background is $u^{\mu}$. Then, the contraction of the four-velocity is $g_{ab}u^au^b= -k$,  with $k=1 $ for timelike events and $ k=0 $ for lightlike events.

The specific energy and specific angular momentum, thus, take the following forms:
\begin{equation}\nonumber
   \mathcal{E}=-u^{\mu}\psi_{\mu} =-g_{\mu \nu}u^{\mu}u^{\nu} \ \quad \text{and} \ \quad  \mathcal{L}=u^{\mu}\chi_{\mu}.
\end{equation}
These tensorial equations help us evaluate $\dot{t}$ and $\dot{\phi}$, these are
\begin{equation}\label{54}
\dot{t}={r^2 \over \Delta}\mathcal{E} \ \ \ \text{and} \ \ \ \ \ \dot{\phi} = {\mathcal{L} \over r^2}
\end{equation}
\par
The radial dynamical variable equation of the equatorial sub-manifold is
\begin{equation}\label{55}
g_{tt}u^{t}u^{t}+g_{rr}u^{r}u^{r}+g_{\theta\theta}u^{\theta}u^{\theta}+g_{\phi\phi}u^{\phi}u^{\phi}=-k.
\end{equation}
Now using equations \eqref{54} in equation \eqref{55}, we get the quadrature of the Reissner-Nordström black hole in the equatorial sub-manifold as
\begin{equation}\label{4.6}
{1 \over 2}\dot{r}^2+{1 \over 2}\bigg(k+ {\mathcal{L}^2 \over r^2}\bigg) \bigg[1-{2M \over r}+ {e^2 \over r^2} \bigg]={1 \over 2}\mathcal{E}^2.
\end{equation}
The relativistically effective potential as a function of $\mathcal{L}$, $r$ and $e$ is
\begin{equation}\label{4.7}
V_{\it{eff}}= {1 \over 2}\bigg(k+ {\mathcal{L}^2 \over r^2}\bigg) \bigg[1-{2M \over r}+ {e^2 \over r^2} \bigg].
\end{equation}\par
With no surprise, if we set $e=0$ then $V_{\it{eff}}$ reduces to equation 6.3.14 of \cite{Wald1984} and if we put $k=0$ it reduces to equation 6.3.29 of Wald \cite{Wald1984}.\par
\begin{figure}[h]
\centering \includegraphics[scale=0.65]{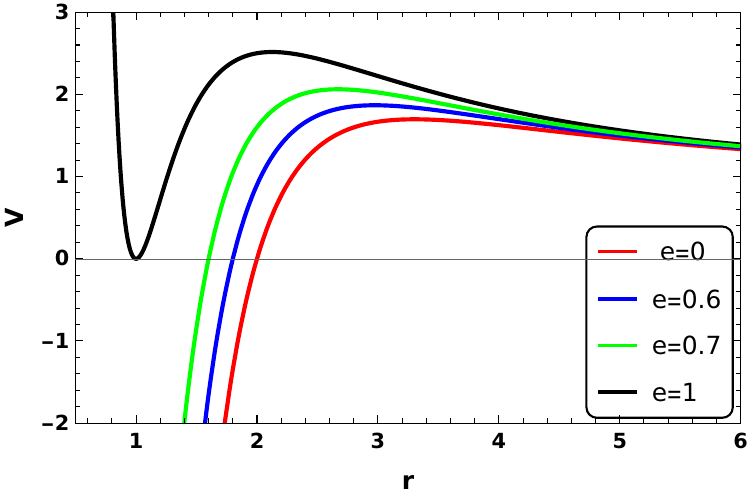}
\caption{\small Variation of effective potential of Reissner-Nordström blackhole with $r$ for different values of $e$.}
\label{variatio}
\end{figure}

The graphical representation of the effective potential of the Reissner-Nordström black hole w.r.t. $r$ for different values of charge $e$ is as shown in Figure \ref{variatio}. We see that the radius of circular orbits decreases with increase in black hole charge for the same effective potential.
For a test particle to perform circular motion, both  $V_{\it{eff}}$ and its first derivative  with respect to $ r $ must be zero. That is,
\begin{equation}\nonumber
V'_{\it{eff}}={kM \over r^2}-{ke^2 \over r^3}-{\mathcal{L}^2 \over r^3}+{3\mathcal{L}^2 M \over r^4}-{2e^2\mathcal{L}^2 \over r^5}=0
\end{equation}
and the roots of this equation are
\begin{eqnarray}\label{4.8} 
r&=& \frac{\mathcal{L}^2+e^2}{3 M}+{\big\{P+\sqrt{Q}\big\} \over 3 (2)^{1 \over 3} M} \nonumber \\&+&\frac{2^{1 \over 3}\big\{9 \mathcal{L}^2 M^2-(\mathcal{L}^2+e^2)^2\big\}}{3M\big\{P+\sqrt{Q}\big\}},
\end{eqnarray}
here we have defined new functions $P$ and $Q$, respectively, as
\begin{eqnarray}
P=\Big[2\mathcal{L}^6-27 \mathcal{L}^4 M^2+6 \mathcal{L}^4 e^2&+&27 \mathcal{L}^2 M^2 e^2+ \nonumber \\ &+&6 \mathcal{L}^2 e^4 +2e^6\Big]
\end{eqnarray}
and
\begin{eqnarray}
 Q&=&\bigg[\Big\{2\mathcal{L}^6-27\mathcal{L}^4 M^2+6\mathcal{L}^4 e^2+27\mathcal{L}^2 M^2 e^2 \nonumber \\& +&6\mathcal{L}^2 e^4 +2e^6\Big\}^2 +4\Big\{9\mathcal{L}^2 M^2-(\mathcal{L}^2+e^2)^2\Big\}^3\bigg]^{1 \over 3} \quad
\end{eqnarray}
The sufficient and necessary condition to be satisfied by $Q$ is $M^2 > e^2$. For a particle to execute circular motion in the Reissner-Nordstrom field, the conditions to be satisfied are
\begin{equation}\nonumber
\mathcal{E}^2 = V_{\it{eff}}(r) \qquad \text{and} \qquad  V'_{\it{eff}}(r)=0.
\end{equation}
For a timelike geodesic ($k=1$), the energy for a massive particles is
\begin{equation}\label{4.10}
{\mathcal{E}^2 \over 2}= {1 \over 2}\bigg(1+ {\mathcal{L}^2 \over r^2}\bigg) \bigg[1-{2M \over r}+ {e^2 \over r^2} \bigg]=V_{\it{eff}}.
\end{equation}
From $V'_{\it{eff}}(r)=0$ and using $\mathcal{L}^2$ in equation \eqref{4.10}, we can easily check 
\begin{equation}\label{4.11}
\mathcal{L}=\sqrt{{Mr^3-e^2r^2 \over r^2-3Mr+2e^2}}
\end{equation}
and
\begin{equation}\label{4.11a}
\mathcal{E}=\sqrt{{1 \over r^2} \ {(r^2-2Mr+e^2)^2 \over r^2-3Mr+2e^2}}.
\end{equation}
Reduction of Reissner-Nordström metric to Schwarzschild metric can perfectly convert equations \eqref{4.11} and \eqref{4.11a} into their Schwarzschild counterparts, see \cite{AlZehrani2021}. The condition for Reissner-Nordström black hole to have ISCO necessitates that the second derivative of $V_{\it{eff}}(r)$ should also vanish, i.e.,
\begin{equation}\label{4.12}
-{2M \over r^3}+{3e^2 \over r^4}+{3\mathcal{L}^2 \over r^4}-{12\mathcal{L}^2M \over r^5}+{10e^2\mathcal{L}^2 \over r^6}=0.
\end{equation}
The roots of equation \eqref{4.12} that we call $r_{\mathcal{ISCO}}$, with condition $M^2\geqslant e^2$, were found as
\begin{eqnarray}\label{4.13}
r_{\mathcal{ISCO}}&=&2M+\frac{4 M^3-3 M e^2}{\big\{8 M^6-9 M^4 e^2+2 M^2 e^4+\sqrt{W}\big\}^{1 \over 3}}\nonumber \\ &+& \frac{\big\{8 M^6-9 M^4 e^2+2 M^2 e^4+\sqrt{W}\big\}^{1 \over 3}}{M}
\end{eqnarray}
where we have defined a new function,
\begin{equation}
    W=5 M^8 e^4-9 M^6 e^6+4 M^4 e^8.
\end{equation}
We can find the specific energy and the specific angular momentum in the ISCO region by invoking equation \eqref{4.13} in equations \eqref{4.11} and \eqref{4.11a}. However, skipping writing the explicit forms due to their complex and tedious forms, we will keep them in terms of $r_{\mathcal{ISCO}}$ as
\begin{equation}
\mathcal{L_{ISCO}}=\sqrt{{Mr_{\mathcal{ISCO}}^3-e^2r_{\mathcal{ISCO}}^2 \over r_{\mathcal{ISCO}}^2-3Mr_{\mathcal{ISCO}}+2e^2}}
\end{equation}
\begin{equation}
\mathcal{E_{ISCO}}=\sqrt{{1 \over r_{\mathcal{ISCO}}^2} \ {(r_{\mathcal{ISCO}}^2-2Mr_{\mathcal{ISCO}}+e^2)^2 \over r_{\mathcal{ISCO}}^2-3Mr_{\mathcal{ISCO}}+2e^2}}.
\end{equation}
\subsection{ISCO of charged particles in Reissner-Nordström black hole}\label{iscornbh}
After discussing the relativistic kinematic properties of neutral particles in the Reissner-Nordstrom field, we are in a position to discuss the electrodynamic properties of charged particles in the Reissner-Nordstrom field. As already stated, the Reissner-Nordström metric allows two symmetries $\textemdash$ temporal and azimuthal. We can use these two symmetries to construct the relativistic electromagnetic four potential.
\begin{figure}[h]
\includegraphics[scale=0.65]{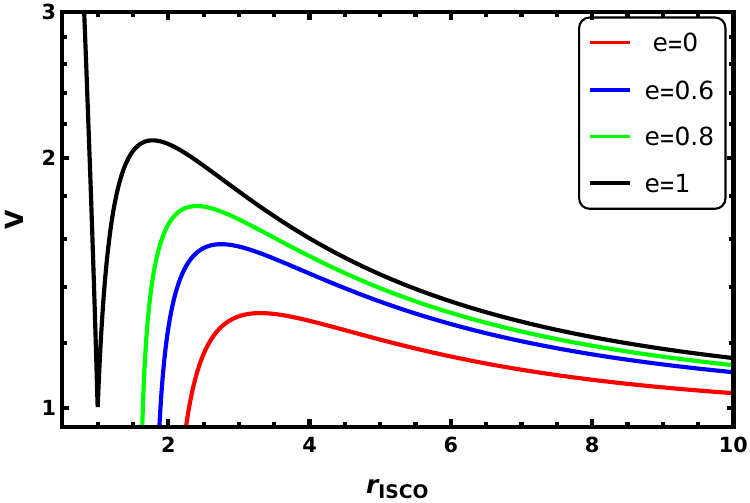}
\includegraphics[scale=0.47]{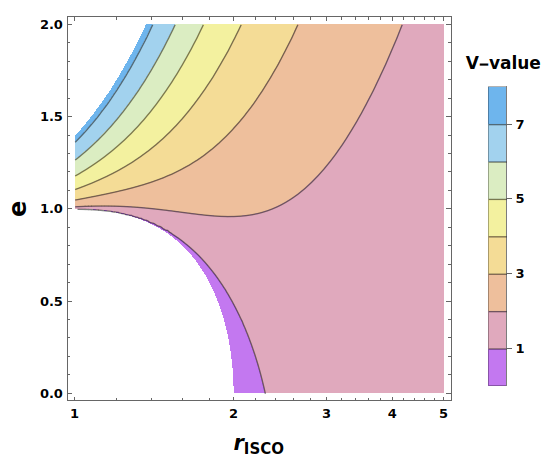}
\caption{\small Variation of $r_{\mathcal{ISCO}}$ with potential for different values of $e$ (top) and charge of black hole $e$ for different potential depths (bottom).}
\label{ar}
\end{figure}
Therefore, by ansatz
\begin{equation}
A^{\mu}={e \over r}-{Q \over 2M}\psi_{(t)}^{\ \mu}+{B \over 2}\psi_{(\phi)}^{\ \mu},
\end{equation}
where $Q$ is the charge of the black hole and $B$ is the axisymmetric magnetic field of the black hole. However, it must be noted that the Reissner-Nordström field has it's own intrinsic charge and is denoted by $e$. Analogous to Schwarzschild and Kerr field we can find the radial dynamical equation in an equatorial sub-manifold as
\begin{eqnarray}
\dot{r}&=&\Bigg({\mathcal{E}^2-1 \over 1-\mathcal{E}a}\Bigg)-\Bigg[-{1 \over r}+{\mathcal{L} \over r^2}-{a^2-e^2 \over r^2(1-\mathcal{E}a)} \nonumber \\ &-&\Bigg(1-{e^2 \over 2r}\Bigg)\Bigg(\mathcal{L}^2-{a^2-e^2 \over r^2(1-\mathcal{E}a)} +{a^2 -e^2 \over 1-\mathcal{E}a}\Bigg){1 \over r^3}\Bigg] \nonumber,
\end{eqnarray}
where, $a$ is a charge parameter defined as $eQ$.\par
We can simplify above equation by factorizing the quadrature as
\begin{equation}
\dot{r}={r^4 \over \mathcal{L}^2}\big(\mathcal{E}-V_{+}\big)\big(\mathcal{E}-V_{-}\big).
\end{equation}
From this equation we can find the effective potential, by still confining our discussion to the weak field approximations, as
\begin{equation}
V_{\pm}={e \over r} \pm \sqrt{1-{2 \over r}+{\mathcal{L}^2 \over r^2}-{2\mathcal{L}^2 \over r^3}+{e^2 \over r^2}+{\mathcal{L}^2e^2 \over r^4}}.
\end{equation}
An easy and logical comprehension of this equation is possible by the graphical interpretation of variation of $r_{\mathcal{ISCO}}$ with potential for different values of $e$ and charge of black hole $e$ for different potential depths as shown in figure \ref{ar}.

Now, by an already followed procedure, in order to find the ISCO, which is located at an inflection manifold, the first derivative and also the second derivative of effective potential ought to vanish. With these conditions, we found the necessary condition for ISCO of the charged particle in the Reissner-Nordström spacetimes as
\begin{widetext}
\begin{eqnarray}\nonumber
\mathcal{E}\big\{3e^2-2r_{\mathcal{ISCO}}\big\}\sqrt{9\mathcal{E}^2r_{\mathcal{ISCO}}^4-4(5\mathcal{E}^2+4)r_{\mathcal{ISCO}}^3+12(\mathcal{E}^2+3)r_{\mathcal{ISCO}}^2-24r_{\mathcal{ISCO}}+4e^2} \nonumber \\ \nonumber \\ + \ 6\mathcal{E}^2r_{\mathcal{ISCO}}^3-3(2+(3e^2+2)\mathcal{E}^2)r_{\mathcal{ISCO}}^2 +(5\mathcal{E}^2e^2+4e^2+2)r_{\mathcal{ISCO}}-6e^2=0. \qquad
\end{eqnarray}
\end{widetext}
From the plot (Figure \ref{ar}) it is conspicuously clear that charged black holes enhance the stability of circular orbits, check by comparing potentials for different values of charge of black hole.\par
The value of $a$ for which the minimal ISCO exists is given by the relation
\begin{equation}\label{aeiou}
a=2e\sqrt{{-5e^2+9\big(1-\sqrt{1-e^2}\big) \over 25e^2-9}}.
\end{equation}
The above equation shows the electromagnetic coupling effects in the vicinity of strong gravitational fields.
The highest value of $r_{\mathcal{ISCO}}$ for which ISCO solutions can be found when $e=1$ is
\begin{equation}\nonumber
r_{\mathcal{ISCO}}=3M.
\end{equation}

    \section{Kerr-Newman Black Hole}\label{knbh}
The next important class of black holes is the famous Kerr-Newman black hole. These are the spherically symmetric rotating charged black holes. The spin parameter for these black holes does not vanish, i.e., $a\neq0$ and let the charge of the Kerr-Newman black hole be $e$. The geometry of Kerr-Newman suggests that the true physical singularity has a ring structure. The metric of the Kerr-Newman black hole in the spherical polar coordinates has a complex and tedious structure. In the Boyer-Lindquist coordinates:
\begin{eqnarray}
\Sigma &=&r^2 +a^2\cos^2\theta, \nonumber \\ \Delta&=& r^2+a^2+e^2-2Mr  \nonumber \qquad \text{and} \\ \Gamma&=&\left(a^2+r^2\right)^2-\Delta a^2\sin^2\theta  \nonumber, 
\end{eqnarray}
the squared line element of the Kerr-Newman metric \cite{Newman1965}, then, is:
\begin{eqnarray}
ds^2&=&\frac{a^2 \sin ^2\theta -\Delta }{\Sigma }dt^2 -2a\frac{(2Mr-e^2)\sin ^2\theta }{\Sigma }dtd\phi \nonumber \\ &+& \frac{\Gamma}{\Sigma }\sin^2\theta d\phi^2 +\frac{\Sigma }{\Delta }dr^2 +\Sigma d\theta^2.
\end{eqnarray}
The Kerr-Newman metric is the most general black hole metric as predicted by the "no-hair" conjecture.
\subsection{Circular orbits of gravitating neutral particles in Kerr-Newman black hole}\label{cogknbh}
In this section circular orbits in the equatorial sub-manifold of Kerr-Newman black hole are evaluated. In the equatorial manifold $ \theta= \frac{\pi}{2}$ and $\dot{\theta}=0.$ The circular orbits in the equatorial sub-manifold are interesting and very insightful.\par
The Kerr-Newmann metric, like other metrics, due to its temporal and azimuthal symmetry, allows two killing vectors
\begin{equation}
\psi^{\mu} = \bigg(\frac{\partial}{\partial t}\bigg)^{\mu} \quad \text{and} \quad \chi^{\mu} = \bigg(\frac{\partial}{\partial \phi}\bigg)^{\mu}.
\end{equation}
As Noether's theorem necessitates two conservation laws corresponding to each symmetry. Corresponding to temporal symmetry is the conservation of specific energy $\mathcal{E}$ and corresponding to azimuthal symmetry is the conservation of specific angular momentum $\mathcal{L}$. The respective quantities are:
\begin{eqnarray}\nonumber
    \mathcal{E} &=&-u^{\mu}\zeta_{\mu}=-\frac{a^2 \sin ^2\theta -\Delta }{\Sigma }\dot{t}   + 2\Bigg\{\frac{a \sin ^2\theta \left(2Mr-e^2\right)}{\Sigma }\Bigg\}\dot{\phi}\\
\end{eqnarray}
and
\begin{eqnarray}
    \mathcal{L}&=&u^{\nu}\psi_{\nu}=2\Bigg\{\frac{a \sin ^2\theta \left(2Mr-e^2\right)}{\Sigma }\Bigg\}\dot{t} \nonumber \\ &+& \sin^2\theta\bigg[\frac{\sin ^2\theta  \Big\{\left(a^2+r^2\right)^2-a^2 \Delta  \sin ^2\theta \Big\}}{\Sigma }\bigg]\dot{\phi}
\end{eqnarray}
\par
If the four-velocity of a test particle in the gravitational background is $u^{\mu} $, then the contraction of the four-velocity is
\begin{equation}\label{kwas}
g_{\mu\nu}u^{\mu}u^{\nu}= -k,
\end{equation}
with $k=1 $ for timelike events and $k=0 $ for lightlike events. The explicit form of equation \eqref{kwas} in the equatorial sub-manifold is
\begin{widetext}
\begin{equation}\label{SSSSS}
\bigg({2Mr-r^2-e^2 \over r^2}\bigg)\dot{t}^2 + {2a \over r^2}\Big(e^2-2Mr\Big)\dot{t}\dot{\phi}+\bigg({r^2 \over r^2+a^2+e^2-2Mr}\bigg)\dot{r}^2+\bigg({r^4+a^2r^2-a^2e^2+2Ma^2r \over r^2}\bigg)\dot{\phi}^2=-k.
\end{equation}
\end{widetext}
Specific energy ($\mathcal{E}) $ and specific angular momentum ($ \mathcal{L} $) in the equatorial sub-manifold of Kerr-Newman spacetime take the forms as
\begin{equation}\label{sa}
    \mathcal{E}=\frac{r^2-e^2-2Mr}{r^2}\dot{t}+ a\frac{2Mr-e^2}{r^2}\dot{\phi}
\end{equation}
and
\begin{equation}\label{sad}
    \mathcal{L}= a\frac{e^2-2Mr}{r^2}\dot{t} + \frac{(r^2+a^2)^2-\Delta a^2}{r^2} \dot{\phi}.
\end{equation}
In order to obtain the radial dynamical variable, equations \eqref{sa} and \eqref{sad} are used to obtain the expressions for $ \dot{t}$ and $\dot{\phi} $ in terms of $\mathcal{E}$ and $\mathcal{L}$, and are, therefore, expressed  respectively as .
\begin{figure}[h]
\includegraphics[scale=0.8]{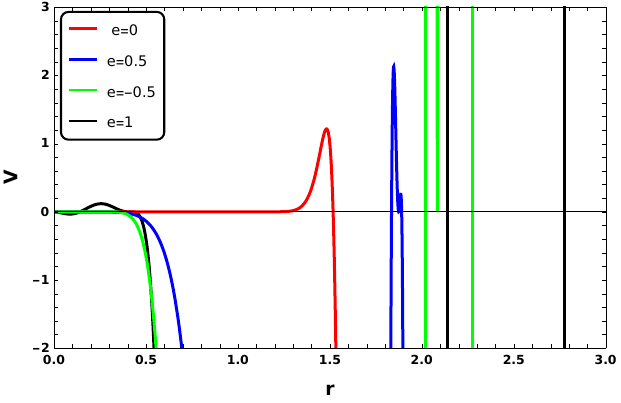} \ \ \includegraphics[scale=0.9]{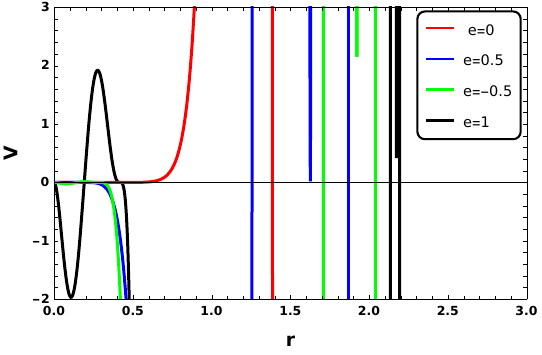}
\caption {\small Variation of effective potential of Kerr-Newman blackhole with $r$ for different values of $e$ with $\mathcal{E}=1$ , $a=1$ (top) and $\mathcal{E}=0.8$ , $a=1.5$ (bottom).}
\label{variation}
\end{figure}

\begin{figure}[H]
   \centering
   \includegraphics[scale=0.7]{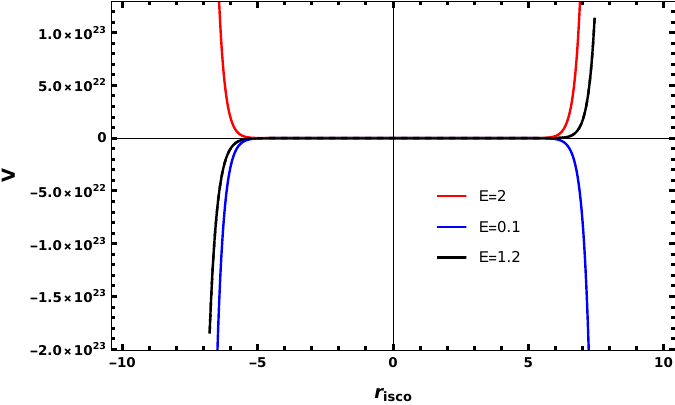} 
   \caption{\small The variation of potential w.r.t. $r_{\it{ISCO}}$ for different values of energy $\mathcal{E}$}
   \label{quanta}
\end{figure}

\begin{equation}\nonumber
\dot{t}=\frac{r^2(a\mathcal{L}(e^2-2Mr)+r^4\mathcal{E}+a^2\mathcal{E}(-e^2+r(2M+r))}{a^2 \left(-2e^2 r (2 M+r)+2 e^4+r^4\right)+r^4 \left(r (r-2 M)-e^2\right)} 
\end{equation}
and
\begin{equation}\nonumber
\dot{\phi}=\frac{r^2 \left(-a \mathcal{E} \left(e^2-2 M r\right)+\mathcal{L} r (r-2 M)-\mathcal{L} e^2\right)}{a^2 \left(-2 e^2 r (2 M+r)+2 e^4+r^4\right)+r^4 \left(r (r-2 M)-e^2\right)}
\end{equation}

Now using these equations in equation \eqref{SSSSS}, we get the radial dynamical equation as
\newpage
\begin{widetext}
\begin{eqnarray}\label{Bob}\nonumber
\dot{r}^2 &=& \Bigg\{ -k + \bigg[{r^2+e^2-2Mr \over r^2}\bigg]\bigg[\frac{r^2(a\mathcal{L}(e^2-2Mr)+r^4\mathcal{E}+a^2\mathcal{E}(-e^2+r(2M+r))}{a^2 \left(-2e^2 r (2 M+r)+2 e^4+r^4\right)+r^4 \left(r (r-2 M)-e^2\right)}\bigg]^2 + {2a \over r^2} \Big(2Mr-e^2\Big) \times \\\nonumber &\times& \bigg[\frac{r^2(a\mathcal{L}(e^2-2Mr)+r^4\mathcal{E}+a^2\mathcal{E}(-e^2+r(2M+r))}{a^2 \left(-2e^2 r (2 M+r)+2 e^4+r^4\right)+r^4 \left(r (r-2 M)-e^2\right)}\bigg]\bigg[\frac{r^2 \left(-a \mathcal{E} \left(e^2-2 M r\right)+\mathcal{L} r (r-2 M)-\mathcal{L} e^2\right)}{a^2 \left(-2 e^2 r (2 M+r)+2 e^4+r^4\right)+r^4 \left(r (r-2 M)-e^2\right)}\bigg]\\ &-& \bigg[{r^4+a^2r^2-a^2e^2+2Ma^2r \over r^2}\bigg]\bigg[\frac{r^2 \left(-a \mathcal{E} \left(e^2-2 M r\right)+\mathcal{L} r (r-2 M)-\mathcal{L} e^2\right)}{a^2 \left(-2 e^2 r (2 M+r)+2 e^4+r^4\right)+r^4 \left(r (r-2 M)-e^2\right)}\bigg]^2\Bigg\}\Bigg/\bigg[{r^2 \over r^2+a^2+e^2-2Mr}\bigg]. \qquad
\end{eqnarray}
\end{widetext}
Verification of this radial equation can be carried out by using the fact that when the Kerr parameter $a=0$ the Kerr-Newman metric reduces to the Reissner-Nordström metric. Therefore,  this reparametrization for timelike geodesics $(k=1)$ reduces \eqref{Bob} into 
\begin{equation}
{1 \over 2}\dot{r}^2+{1 \over 2}\bigg(k+ {\mathcal{L}^2 \over r^2}\bigg) \bigg[1-{2M \over r}+ {e^2 \over r^2} \bigg]={1 \over 2}\mathcal{E}^2,
\end{equation}
which is exactly the quadrature of the Reissner-Nordström black hole, that is, equation \eqref{4.6}, as expected.
In the same analogy, a relatively difficult test is if we set $e=0$ and $a\neq0$ then equation \eqref{Bob} turns into equation \eqref{43}, as should be. If we set both $a=0$ and $e=0$ which means the blackhole under consideration is charge-less and non-rotating. This parametrization converts equation \eqref{Bob} into equation 7.163 of \cite{TPad}, which is a Schwarzschild black hole, as it ought to be. Hence, equation \eqref{Bob} is the most general radial dynamical equation of a particle in the equatorial sub-manifold around a black hole. Equation \eqref{Bob} also assumes two solutions, i.e., positive and negative solutions. After a lot of algebraic computation, the analytic solution of this equation for timelike geodesics obtained is
\begin{widetext}
\begin{equation}
\dot{r}_{\pm}=\pm \frac{\sqrt{\begin{array}{c}
\big\{a^2-2 M r+e^2+r^2\big\}\big\{a^4 (\mathcal{E}^2 r^6 (r (2 M+r)-e^2)-(-2 e^2 r (2 M+r)+2 e^4+r^4)^2)+  \\ +2 a^3 \mathcal{E} \mathcal{L} r^6 (e^2-2 M r)+a^2 r^2 (\mathcal{L}^2 (-4 e^4 r (2 M+r)+r^5 (2 M-r)+4 e^6+3 e^2 r^4)+  \\ +\mathcal{E}^2 r^4 (-4 M^2 r^2+4 M e^2 r-e^4+2 r^4)-2 r^2 (r (r-2 M)-e^2) (-2 e^2 r (2 M+r)+2 e^4+r^4))-  \\ -2 a \mathcal{E} \mathcal{L} r^6 (2 M r-e^2) (r (r-2 M)+e^2)-r^6 (\mathcal{L}^2+r^2) (r (2 M-r)+e^2)^2+\mathcal{L}^2 r^{10} (r (r-2 M)+e^2)\big\}\end{array}}}{r\big\{a^2 (-2 e^2 r (2 M+r)+2 e^4+r^4)+r^4 (r (r-2 M)-e^2)\big\}}
\end{equation}
\end{widetext}
This radial dynamical equation, obviously, is the most general of all the other aforementioned dynamical equations. Now, If we substitute $e=0$, this radial equation takes the form of equation \eqref{45}, as, presumably, it should be.\par
Now, the most important step of this problem is to find the quadrature of Kerr-Newman black holes. We have
\begin{widetext}
\begin{equation}
\dot{r}^2=-\frac{\begin{array}{c}
\big\{a^2+r (r-2 M)+e^2\big\} \big\{a^4 (\mathcal{E}^2 r^6 (e^2-r (2 M+r))+(-2 e^2 r (2 M+r)+2 e^4+r^4)^2\big) +2 a^3 \mathcal{E} \mathcal{L} r^6 (2 M r-e^2)+  \\ +a^2 r^2 (\mathcal{L}^2 (4 e^4 r (2 M+r)+r^5 (r-2 M)-4 e^6-3 e^2 r^4)+\mathcal{E}^2 r^4 (4 M^2 r^2-4 M e^2 r+e^4-2 r^4)+  \\ +2 r^2 (r (r-2 M)-e^2) (-2 e^2 r (2 M+r)+2 e^4+r^4))+2 a \mathcal{E} \mathcal{L} r^6 (2 M r-e^2)(r (r-2 M)+e^2)+  \\ +r^6 (\mathcal{L}^2+r^2) (r (2 M-r)+e^2)^2-(\mathcal{E}^2 r^{10} (r (r-2 M)+e^2))\big\}\end{array}}{\big\{a^2 r \left(-2 e^2 r (2 M+r)+2 e^4+r^4\right)+r^5 \left(r (r-2 M)-e^2\right)\big\}^2}.
\end{equation}
\end{widetext}

So, the relativistic effective potential as a function of $r,\mathcal{L}$ and $ \mathcal{E}$ of a gravitating test particle in the equatorial sub-manifold of the Kerr-Newman field is:
\begin{widetext}
\begin{equation}
V_{\it{eff}}(r,\mathcal{L}, \mathcal{E})=-\frac{\begin{array}{c}
\big\{a^2+r (r-2 M)+e^2\big\} \big\{a^4 (\mathcal{E}^2 r^6 (e^2-r (2 M+r))+(-2 e^2 r (2 M+r)+2 e^4+r^4)^2\big) +2 a^3 \mathcal{E} \mathcal{L} r^6 (2 M r-e^2)+  \\ +a^2 r^2 (\mathcal{L}^2 (4 e^4 r (2 M+r)+r^5 (r-2 M)-4 e^6-3 e^2 r^4)+\mathcal{E}^2 r^4 (4 M^2 r^2-4 M e^2 r+e^4-2 r^4)+  \\ +2 r^2 (r (r-2 M)-e^2) (-2 e^2 r (2 M+r)+2 e^4+r^4))+2 a \mathcal{E} \mathcal{L} r^6 (2 M r-e^2)(r (r-2 M)+e^2)+  \\ +r^6 (\mathcal{L}^2+r^2) (r (2 M-r)+e^2)^2-(\mathcal{E}^2 r^{10} (r (r-2 M)+e^2))\big\}\end{array}}{2\big\{a^2 r \left(-2 e^2 r (2 M+r)+2 e^4+r^4\right)+r^5 \left(r (r-2 M)-e^2\right)\big\}^2}
\end{equation}
\end{widetext}

The graphical representation of the effective potential of the Kerr-Newman spacetime w.r.t. $r$ for different values of charge $e$ is shown in figure \ref{variation}.

For a particle outside an ergosphere, to execute circular motion, both $V_{\it{eff}}$ and its first derivative with respect to $r $ must vanish. Which gives 
\begin{widetext}
\begin{eqnarray}\nonumber
\left[\begin{array}{c}
\Big\{(r (r-2 M)-e^2) r^5+a^2 (2 e^4-2 r (2 M+r) e^2+r^4) r\Big\} \Big\{-\big\{(2 r-2 M) ((r (r-2 M)-e^2) r^5+a^2 \\ (2 e^4  -2 r (2 M+r) e^2+ r^4) r) (-((e^2+r (r-2 M)) \mathcal{E}^2 r^{10}) +(e^2+(2 M-r) r)^2 \\ (\mathcal{L}^2+r^2) r^6+2 a^3 \mathcal{L} (2 M r-e^2) \mathcal{E} r^6+ 2 a \mathcal{L} (2 M r-e^2) (e^2+r (r-2 M)) \mathcal{E} r^6+a^2 ((e^4-4 M r e^2- \\ 2 r^4+4 M^2 r^2) \mathcal{E}^2 r^4+2 (r (r-2 M)-e^2) (2 e^4-2 r (2 M+r) e^2+r^4) r^2 +\mathcal{L}^2 (-4 e^6 \\ +4 r (2 M+r) e^4  - 3 r^4 e^2+r^5 (r-2 M)))  r^2+a^4 ((e^2-r (2 M+r)) \mathcal{E}^2 r^6+ \\ (2 e^4-2 r (2 M+r) e^2+r^4)^2))\big\} -2 (a^2+e^2+ r (r-2 M)) ((r (7 r-12 M)-5 e^2) r^4+a^2 (2 e^4-2 r \\ (4 M+3 r) e^2+5 r^4)) (-((e^2+r (r-2 M)) \mathcal{E}^2 r^{10})+(e^2+(2 M-r) r)^2 (\mathcal{L}^2+r^2) r^6+2 a^3 \\ \mathcal{L} (2 M r-e^2) \mathcal{E} r^6+2 a \mathcal{L} (2 M r-e^2) (e^2+r (r-2 M)) \mathcal{E} r^6+ a^2 ((e^4-4 M r e^2-2 r^4+4 M^2 r^2) \mathcal{E}^2 r^4+ \\ 2 (r (r-2 M)-e^2) (2 e^4-2 r (2 M+r) e^2+r^4) r^2+\mathcal{L}^2 (-4 e^6+4 r (2 M+r) e^4 -3 r^4 e^2 +r^5 (r-2 M))) r^2  +a^4 \\ ((e^2-r (2 M+r)) \mathcal{E}^2 r^6+(2 e^4-2 r (2 M+r) e^2+r^4)^2))+2 (a^2+e^2+r (r-2 M)) ((r (r-2 M)-e^2) \\ r^5+a^2 (2 e^4-2 r (2 M+r) e^2+r^4) r) \big\{(5 e^2+r (6 r-11 M)) \mathcal{E}^2 r^9 \\ +2 a^3 \mathcal{L} (3 e^2-7 M r) \mathcal{E} r^5+ 2 a \mathcal{L} (3 e^4+2 r (2 r-7 M) e^2+M r^2 (16 M-9 r)) \mathcal{E} r^5 \\ +a^2 (8 r^2 e^6-3 r^4 (\mathcal{E}^2+8) e^4 +2 r^4 (-24 M^2+7 r \mathcal{E}^2 M+12 r^2) e^2+\mathcal{L}^2 (4 e^6-4 r (3 M+2 r) e^4 \\ +9 r^4 e^2+r^5 (7 M-4 r)) +2 e^6 (5 (\mathcal{E}^2-1) r^2+9 M r-8 M^2 E^2)) r -r^5 (e^2+(2 M-r) r) ((3 e^2+r (8 M-5 r)) \mathcal{L}^2 \\ +2 r^2 (2 e^2+r (5 M-3 r))) +a^4 \big(r^5 (r (7 M+4 r)-3 e^2) \mathcal{E}^2-4 (-2 e^4+2 r (2 M+r) e^2-r^4) (M e^2+(e-r) r (e+r))\big)\big\}\Big\}\end{array}\right]=0,
\end{eqnarray}
\end{widetext}
as a necessary condition. We can justify this equation by four exclusive proofs in this paper, check, for example: equations 7.164 of \cite{TPad}, \eqref{tata}, \eqref{4.7} and \eqref{4.8}. The roots of this equation are beyond the scope of this paper.\\ 
\vspace{0ex}
As for other black holes, the evaluation of ISCO for neutral and charged particles in the Kerr-Newman spacetime is difficult to carry out with the present method. The problem has been studied by the Hamilton Jacobi method by others; see \cite{Schroven2017, Sakia2021}. We have, however, tried to study the variation of effective potential of Kerr-Newman black hole w.r.t. $r_{\it{ISCO}}$ for different values of energy $\mathcal{E}$, see figure \ref{quanta}. We also investigated the variation of $r_{ISCO}$ with $V$ for different values of Kerr parameter $a$ and the variation of $r_{ISCO}$ with the black hole charge for Kerr parameter $a=0.1$ in different potential depths, see figure \ref{trees}.
\begin{figure}[h]
    \includegraphics[scale=0.9]{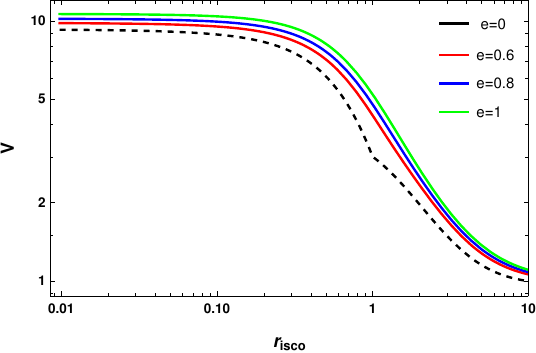} \\ 
\caption{\small The variation of $r_{ISCO}$ with $V$ for different values of Kerr parameter $a$.}
    \label{trees}
\end{figure}
\begin{figure}[h]    
    \includegraphics[scale=0.45]{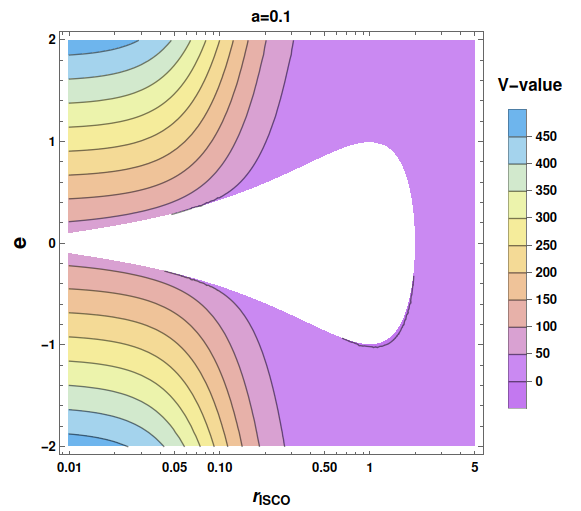}
    \caption{\small The variation of $r_{ISCO}$ with the charge of blackhole for Kerr parameter $a=0.1$ in different potential depths.}
    \label{trees}
\end{figure}
\vspace{-1ex}
    \section{Conclusion}
We have studied the motion of particles, both neutral and charged, around different black holes. Building upon the work already accomplished for Schwarzschild black hole. We started with the dynamics of Kerr black hole followed by the Reissner-Nordström black hole and the  Kerr-Newman black hole. For the Schwarzschild black hole, we know that the closest distance that a particle can revolve around the black hole without falling into it is $6M$ and it loses $5.7\% $ of its rest mass energy before finally spiraling into the ISCO \textemdash as the existence of ISCO around a densely compact object is a purely relativistic phenomenon, therefore, any intuitional conclusion one might have about its behavior turns out to be invalid. This also suggests that, indeed, by gravitational radiative processes, in the astrophysically plausible phenomenon, large amounts of energy, as much as $42\%$, can be radiated. The radius of ISCO of Schwarzschild black hole in electric and magnetic fields, whenever the magnitude of charge ($|q|$) $>0$, increases and the following effects of the magnetic field on radius ($r_{\mathcal{ISCO}}$): it brings the radius of ISCO closer to the coordinate singularity, it makes the maximum value of charge always greater than $M$ for which $r_{\mathcal{ISCO}}$ exists. \par After extending the same method to Kerr black holes, a few interesting results emerged: the $r_{\mathcal{ISCO}}$, for neutral particles, lies in the interval $[M,9 M]$. A particle entering the ISCO region of a Kerr spacetime will release an energy fraction, up-to $19\%$, of its rest mass energy, before finally spiraling into this special orbit. For charged particles orbiting the Kerr spacetime, we saw that an increase in the coupling of black hole and test particle charge sharpens the ISCO region and stabilizes the angular momentum. We checked by putting $a=0$ in the dynamical equations of the Kerr black hole that they, as expected, reduces the Kerr spacetime solution to the Schwarzschild one. Evidently, if $(a^2\mathcal{E}^2-a^2-\mathcal{L}^2)^2 < 12 M^2(\mathcal{L}-a\mathcal{E})^2$ then there is no maxima or minima of $V(r,\mathcal{E}, \mathcal{L})$ which means the particle is trapped and the information is lost. For future research it might be interesting to deduce the same result by using the Heisenberg's uncertainty principle for particles in strong gravitational fields. A very fascinating result emerged from our calculations that is, when $q=M$ a charged particle can be entangled to a Kerr black hole even if they are infinitely apart. \par However in our calculations, we have used the weak field approximation, but the charges have the potential to produce curvature comparable to the curvature produced by the black holes' mass near the event horizon, where the weak field approximations likely breakdown. \par Similar calculations for Reissner-Nordström black holes, with the condition: $M^2 \geq e^2$, yielded that the highest value of $r_{\mathcal{ISCO}}$ for which ISCO solutions can be found when $e = 1$ is $3 M$. The size and existence of ISCO for charged particles depend explicitly on how much the spacetime curvature is being tweaked by the black hole charge $Q$ and the coupling product. So, it is conspicuously clear that the charged black holes enhance the stability of circular orbits. \par In case of the Kerr-Newman black holes, we again followed the same method, we verified our results by different tests, such as: if we set $e=0$ and $a\neq0$, solutions reduced to that of Kerr black hole and if we set $a=0$ and $e=0$, that is a charge-less and non-rotating spacetime, it converts the Kerr-Newman into Schwarzschild black hole. We obtained an expression for effective potential for Kerr-Newman black hole, the expression did not contain magnetic parameter $b$. In future, we would like to create an electromagnetic four potential containing $b$ as well.\par We have altogether overlooked an interesting phenomenon called Outermost Stable Circular Orbits (OSCOs) of black holes due to the computational difficulties. For the future research it may be interesting to study the OSCOs of Kerr-Newman black holes. The study of gravitational wave-shift and Doppler shift at the ISCO and OSCO locations around Black holes can be the promising starting point for future research.
    \begin{acknowledgments}
I inscribe this humble effort of mine, with love and respect, to my Abu \textemdash \ Nazir Ahmad Najar.
\end{acknowledgments}

	\bibliography{References/refs}
	
\end{document}